%% file: muscall.tex
\documentclass{article}
\usepackage[T1]{fontenc} 
\usepackage[utf8]{inputenc} 
\usepackage{ismir,amsmath,cite,url}
\usepackage{graphicx}
\usepackage{color}

\usepackage{times}
\usepackage{epsfig}
\usepackage{graphicx}
\usepackage{amssymb,amsthm,amsmath}

\usepackage[title]{appendix}

\usepackage{microtype}
\usepackage{booktabs} 

\usepackage{url}            
\usepackage{tabularx}       
\usepackage{multirow}
\usepackage{amsfonts}       
\usepackage{nicefrac}       
\usepackage{epigraph}
\usepackage{soul}
\usepackage[dvipsnames, table]{xcolor}
\usepackage{paralist}
\usepackage{caption}
\usepackage{subcaption}
\usepackage{color, colortbl}

\usepackage{pifont} 
\usepackage{wrapfig}

\usepackage{makecell}
\usepackage{array}
\usepackage[export]{adjustbox}

\usepackage{framed}
\usepackage{mathtools}
\usepackage{verbatim} 
\usepackage{enumerate}
\usepackage{bbm}
\usepackage{amsbsy}
\usepackage{float}
\usepackage{enumitem}

\usepackage{lineno}

\usepackage{soul}

\definecolor{mygreen}{rgb}{0.032, 0.6392, 0.2039}
\newcommand{\cmark}{\textcolor{mygreen}{\ding{51}}}%
\newcommand{\xmark}{\textcolor{red}{\ding{55}}}%

\usepackage{booktabs,arydshln}
\makeatletter
\def\adl@drawiv#1#2#3{%
        \hskip.5\tabcolsep
        \xleaders#3{#2.5\@tempdimb #1{1}#2.5\@tempdimb}%
                #2\z@ plus1fil minus1fil\relax
        \hskip.5\tabcolsep}
\newcommand{\cdashlinelr}[1]{%
  \noalign{\vskip\aboverulesep
           \global\let\@dashdrawstore\adl@draw
           \global\let\adl@draw\adl@drawiv}
  \cdashline{#1}
  \noalign{\global\let\adl@draw\@dashdrawstore
           \vskip\belowrulesep}}
\makeatother

\makeatletter
\renewcommand{\paragraph}{%
  \@startsection{paragraph}{4}%
  {\z@}{1.5ex \@plus 1ex \@minus .2ex}{-1em}%
  {\normalfont\normalsize\bfseries}%
}
\makeatother

\def\muscallssl {MusCALL\textsubscript{SSL}}
\def\muscall {MusCALL}
\def\muscallbase {MusCALL\textsubscript{\textsc{Base}}}

\title{Contrastive Audio-Language Learning for Music}

\multauthor
{Ilaria Manco$^{1,2}$ \hspace{1cm} Emmanouil Benetos$^1$ \hspace{1cm} Elio Quinton$^2$ \hspace{1cm} George Fazekas$^1$} {
 $^1$ School of EECS, Queen Mary University of London, London, U.K\\
$^2$ Music \& Audio Machine Learning Lab, Universal Music Group, London, U.K.\\
{\tt\small \{i.manco, emmanouil.benetos, george.fazekas\}@qmul.ac.uk, elio.quinton@umusic.com}
}

\def\authorname{I. Manco, E. Benetos, E. Quinton, G. Fazekas}

\usepackage[bookmarks=false,pdfauthor={\authorname},pdfsubject={\papersubject},hidelinks]{hyperref}

\begin{document}

\maketitle

\begin{abstract}
As one of the most intuitive interfaces known to humans, natural language has the potential to mediate many tasks that involve human-computer interaction, especially in application-focused fields like Music Information Retrieval. In this work, we explore cross-modal learning in an attempt to bridge audio and language in the music domain. To this end, we propose MusCALL, a framework for \underline{Mus}ic \underline{C}ontrastive \underline{A}udio-\underline{L}anguage \underline{L}earning. 
Our approach consists of a dual-encoder architecture that learns the alignment between pairs of music audio and descriptive sentences, producing multimodal embeddings that can be used for text-to-audio and audio-to-text retrieval out-of-the-box. Thanks to this property, MusCALL can be transferred to virtually any task that can be cast as text-based retrieval. 
Our experiments show that our method performs significantly better than the baselines at retrieving audio that matches a textual description and, conversely, text that matches an audio query. We also demonstrate that the multimodal alignment capability of our model can be successfully extended to the zero-shot transfer scenario for genre classification and auto-tagging on two public datasets.
\end{abstract}

\begin{figure}
\centering
 \includegraphics[scale=0.17]{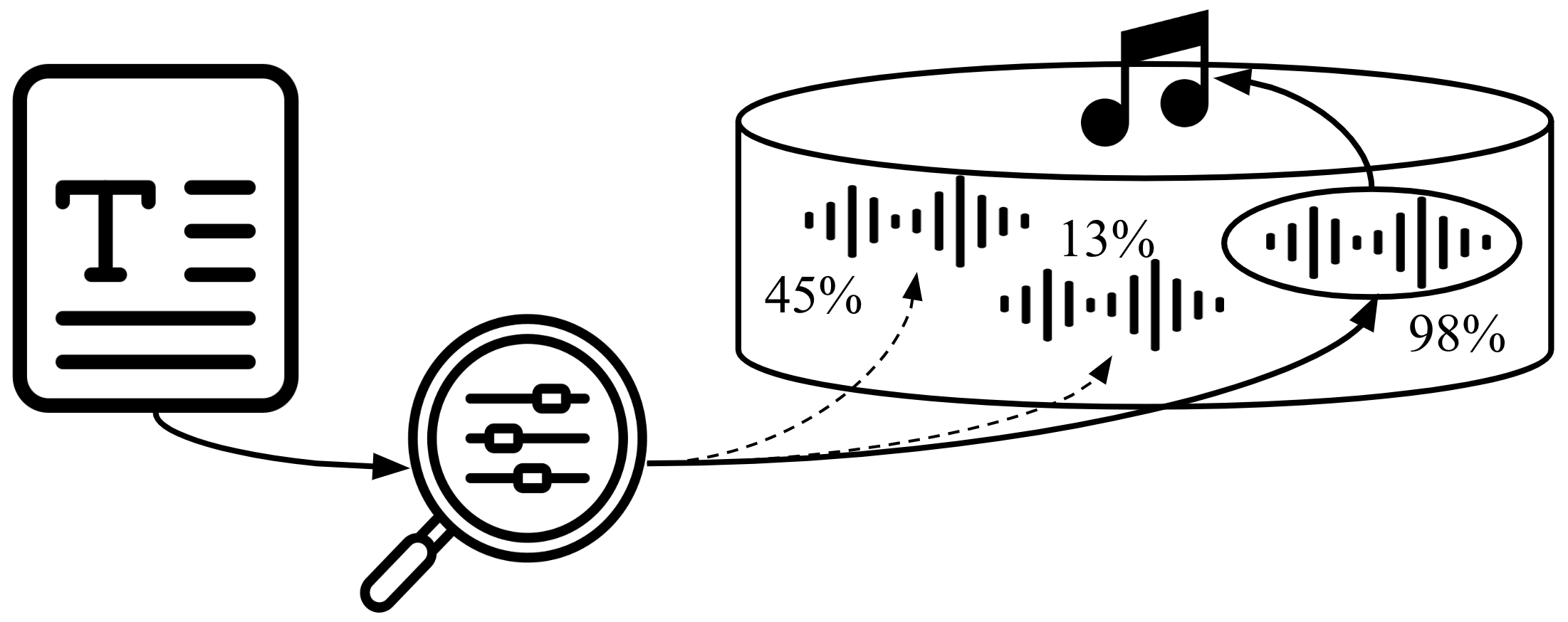}
 \caption{\textbf{Illustration of text-based music retrieval}, where audio items in a database are ranked according to their relevance to a free-form language query.}
 \label{fig:overview}
 \vspace{-0.3cm}
\end{figure}

\section{Introduction}\label{sec:introduction}
Developing effective methods for finding music is at the core of Music Information Retrieval (MIR). Over the years, many approaches have been proposed to browse, search and discover music through a variety of interfaces. Beyond simple search by metadata, existing music retrieval systems allow to express queries via lyrics \cite{Muller2007, Tsukuda2017}, audio examples \cite{Lee2020a}, videos \cite{Li2019a} and humming \cite{Ghias1995}, among others \cite{Muller2019, Knees2020, Watanabe2019}.
Although each of these query types has its merits, none of them supports one of the most popular ways of searching for music today: through free-form text. For example, we commonly look for songs by typing text into a search engine \cite{Hosey2019} or by asking online song naming communities to identify a piece of music we do not have bibliographic information about \cite{Ha2005}.
It becomes evident then that enabling MIR systems to interpret natural language queries can have far-reaching benefits.
This idea is not entirely new to MIR, with prior work \cite{Liem2011, Knees2007, Whitman2002} suggesting similar research directions in the past. So far, however, multimodal systems that integrate natural language have not been widely adopted within the MIR community, possibly due to a lack of suitable datasets or to the practical limitations of NLP methods predating modern language models.

In light of recent breakthroughs in language modelling, we argue that audio-and-language learning has now the potential of closing the semantic gap in MIR \cite{VanDenOord2013}, providing a bridge between computational representations of music signals and the high-level abstractions needed to use those representations in real-world scenarios. 
With this in mind, we propose MusCALL, a method for learning alignment between music-related audio and language data via multimodal contrastive learning. 
Our choice of a contrastive approach is inspired by the recent success of similar methods for joint visio-linguistic modelling (see Section \ref{sec:lang_supervision}). In designing MusCALL, we prioritise the ability to perform retrieval at scale and adopt a dual-encoder architecture, where modalities are processed independently. Compared to multimodal architectures with joint encoders and cross-modal attention mechanisms \cite{Manco2021a}, this design allows to share embedding computations among pairs, resulting in a computationally more efficient model.

With this work, we aim to unify audio and language modelling, paving the way for MIR systems that can interpret language-based queries, as illustrated in Figure \ref{fig:overview}. Our primary contributions can be summarised as follows: (i) we explore multimodal contrastive learning in the context of music-related audio and language for the first time; (ii) we introduce a method for cross-modal retrieval of music, providing the first example of sentence-based music search; (iii) we perform an extensive set of experiments to systematically validate details of our approach and evaluate its performance on popular MIR tasks in a zero-shot setting.\footnote{Code is available at \href{https://github.com/ilaria-manco/muscall}{https://github.com/ilaria-manco/muscall}}

\section{Related Work}\label{sec:related_work}
\subsection{Natural Language Processing in MIR}
Prior works in the MIR literature have explored leveraging natural language in the music domain from different angles. Early efforts focused on text as a modality in isolation, adopting NLP techniques to construct knowledge bases from music-related text corpora \cite{Oramas2016a}, build semantic graphs for artist similarity from biographies \cite{Oramas2015}, or perform genre classification based on album reviews \cite{Oramas2016b}. Recent efforts, more closely related to the present work, have instead started favouring multimodal approaches. These have explored deep learning with multimodal input data, typically audio combined with text such as reviews or lyrics, for applications as varied as music classification and recommendation \cite{Oramas2018}, mood detection \cite{Delbouys2018}, music emotion recognition \cite{Jeon2017} and music captioning \cite{Choi2016, Cai2020, Manco2021}.

\subsection{Audio-Text Cross-modal Learning}
Another related line of work is cross-modal learning that leverages audio and tags as text input.
Works in this area have proposed contrastive learning-based approaches to enrich audio representations via cross-modal alignment, either for general-purpose audio classification \cite{Favory2020, Favory2020a} or for music-focused tasks \cite{Ferraro2021}. Differently from our work, these approaches require finetuning on the downstream tasks and none of them directly uses cross-modal representations. Others have explored pre-trained word embeddings in triplet networks to perform tag-based music retrieval via a multimodal embedding space \cite{Choi2019, Won2020b}.
At a high level, these works share a similar approach to ours and all aim to learn multimodal audio representations by leveraging text. Unlike our work, however, none of them makes use of natural language, using tags as text input instead.

Finally, similarly to our work, \cite{Won2021a} also addresses the problem of matching audio and long-form text for music retrieval, but offers a fundamentally different approach, which relies on bridging the audio and text modalities via a common emotion embedding space.

\subsection{Learning from Language Supervision}\label{sec:lang_supervision}
In the previous sections, we have highlighted some research efforts towards bringing together audio and language, but, ultimately, multimodal learning still occupies a marginal role in MIR and has yet to fully enjoy the benefits of modern language models. 
In adjacent fields such as computer vision, jointly modelling vision and language has instead become a very active area of research, with several successful attempts at using multimodality to develop task-agnostic models that can easily adapt to novel tasks \cite{Lu2019, Chen2020b, Radford2021, Jia2021}. The key insight behind these models is that language captures many of the abstractions humans use to navigate the world and can therefore act as a rich supervisory signal for general-purpose learning, even in tasks that are not directly based on language \cite{Andreas2018}. 
Among these works, CLIP \cite{Radford2021} is a particularly influential example, having demonstrated for the first time that supervision from natural language can induce highly generic visual representations at scale. 

These breakthroughs suggest that natural language supervision has a large potential beyond the image domain. This has recently prompted the adoption of similar approaches in machine listening, where applications to both music \cite{Manco2021a} and non-music audio \cite{Liu2022, Oncescu2021} have started to emerge. A subset of works in this area have proposed to directly extend CLIP by incorporating audio as an additional modality \cite{Wu2021a, Zhao2021, Guzhov2021}. Although similar in spirit to our work, these approaches exploit audio-visual correspondences in video, thus requiring large-scale visio-linguistic pre-training. Like these works, we also borrow from CLIP's contrastive dual-encoder approach, but we adapt this framework to the audio modality without any pre-training, while also addressing the specific set of challenges that arise from joint audio-linguistic modelling in the music domain.

\begin{figure}
 \includegraphics[width=1.0\columnwidth]{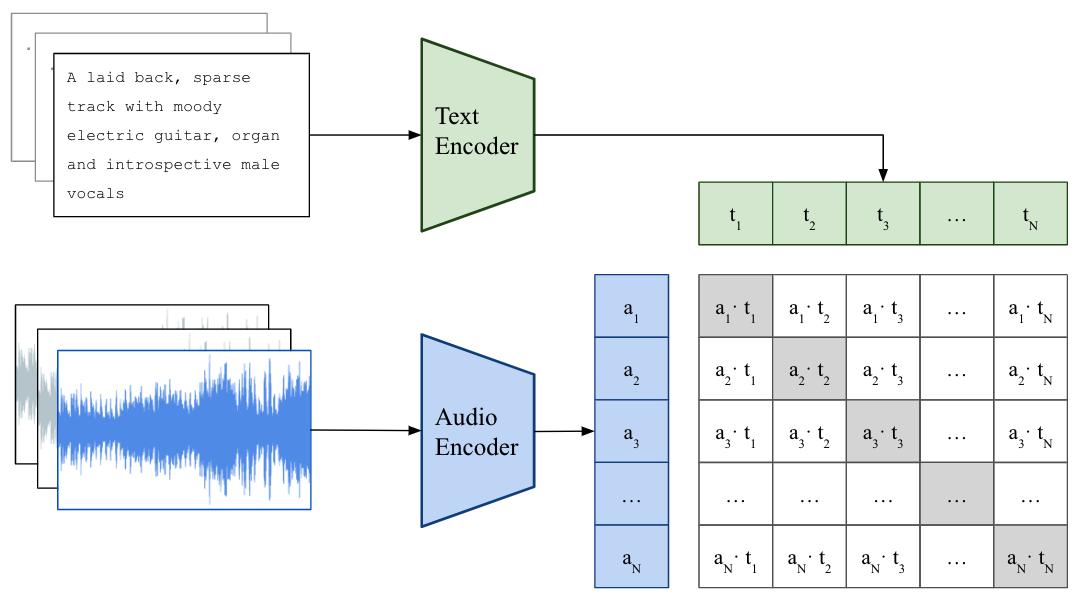}
 \caption{\textbf{Overview of MusCALL}. An audio encoder and a text encoder are trained via a contrastive loss to maximise the similarity between representations of $N$ aligned (audio, text) pairs within a mini-batch. At test time, the similarity between embeddings in the learnt multimodal space is used to rank database items and perform cross-modal retrieval.}
 \label{fig:contrastive_overview}
   \vspace{-0.4cm}
\end{figure}

\section{MusCALL} \label{sec:approach}
\subsection{Multimodal Contrastive Learning (MCL)}\label{sec:mcl}
Contrastive learning has emerged as a powerful paradigm in self-supervised representation learning for images \cite{Chen2020} and audio \cite{Niizumi2021, Saeed2020, Al-Tahan2021, Spijkervet2021}. At their core, all flavours of contrastive learning rely on constructing different views of the data and forcing representations of positive pairs of such views to have, on average, higher similarity than those of negative pairs. This encourages a model to learn representations that are \textit{discriminative} with respect to changes to the content and \textit{invariant} to nuisance transformations. In representation learning, this is a useful property, since it results in more robust and effective representations \cite{Fonseca2020, LeKhac2020, Chen2020a}.
Most formulations of contrastive learning in unimodal scenarios use data augmentations to create multiple views of the inputs. When switching to the multimodal setting, however, different views naturally co-occur in the data, making this a natural testbed for contrastive learning \cite{Alayrac2020, Radford2021, Patrick2020}.

\subsection{Extending MCL to Music and Language}\label{sec:muscall}
In this work, we explore multimodal contrastive learning (MCL) in the context of audio-linguistic data. More specifically, we consider pairs of music audio tracks and their associated textual descriptions and aim to learn a model that is able to predict when items from a given pair match, i.e. when they are semantically aligned. An overview of our approach is shown in Figure \ref{fig:contrastive_overview}. In practice, our goal is to learn two encoders, $f_a(\cdot)$ for the audio modality and $f_t(\cdot)$ for the text modality, such that for any given (audio, text) pair formed by the tuple $(a_i, t_i)$, the resulting L2-normalised embeddings $\boldsymbol{z}_{a, i} = f_a (a_i)$ and $\boldsymbol{z}_{t, i} = f_t (t_i)$ lie closely in the joint embedding space, with respect to some distance metric, only if $a_i$ and $t_i$ represent similar content. Given our problem definition, we achieve this by optimising a type of N-pair contrastive loss, known as InfoNCE \cite{VanDenOord2018}. Based on this, our audio-to-text loss can be defined as follows:
\begin{equation}\label{eq:nce}
    \mathcal{L}_{a \rightarrow t}
    = - \frac{1}{N} \sum_{i}^{N} \log \frac{\exp \left(\boldsymbol{z}_{a, i} \cdot \boldsymbol{z}_{t, i}^{+} / \tau\right)}
    {\sum\limits_{\boldsymbol{z} \in\left\{\boldsymbol{z}_{t, i}^{+}, \boldsymbol{z}_{t, i}^{-}\right\}} 
    \exp \left(\boldsymbol{z}_{a, i} \cdot \boldsymbol{z} / \tau\right)},
\end{equation}
where $N$ is the batch size, $\boldsymbol{z}_{t, i}^{+}$ and $\boldsymbol{z}_{t, i}^{-}$ are embeddings of positive and negative text samples for item $a_i$, and $\tau$ is a temperature parameter used to scale the similarity scores.
By noting that $\mathcal{L}_{t \rightarrow a}$ can be defined symmetrically to Eq. (\ref{eq:nce}), the total loss over all (audio, text) pairs in a mini-batch is simply obtained by summing the two losses together:
\begin{equation}\label{eq:loss_mcl}
    \mathcal{L}_{a \leftrightarrow t} = 
    \mathcal{L}_{a \rightarrow t} + \mathcal{L}_{t \rightarrow a}.
\end{equation}

Given a dataset of tuples $\mathcal{D} = \{(a_i, t_i)\}_{i=1:D}$, there are several plausible strategies for constructing positive and negative pairs.
One such way, and arguably the most intuitive, is to follow the implicit alignment found in the data. In this case, for each sample $i$, the positive and negative sets within a mini-batch become: $\boldsymbol{z}^{+}_{x, i} = \{\boldsymbol{z}_{y, i}\}$ and $\boldsymbol{z}^{-}_{x, i} = \left\{z_{y, j} \mid \forall j \in \{1, \dots, N\}, j \neq i\right\}$, where $x$ and $y$ denote the two different modalities. We refer to this sampling strategy as \textit{instance discrimination} \cite{Saunshi2019}.

\subsubsection{Content-Aware Loss Weighting}\label{sec:lw}
Constructing positive and negative samples by instance discrimination is a design choice that relies on two implicit assumptions: firstly, that all items in the dataset are correctly aligned, i.e. that each (audio, text) pair is formed by the most semantically close items amongst all possible pairings; secondly, that all non-aligned pairs represent sufficiently different content and are therefore equally valid as negatives. In a real-world dataset, this is unlikely to hold true in all cases. Intuitively, within a randomly sampled mini-batch, some tracks will share similarities with other tracks, and so will their respective captions.

The above observations suggest that a more informed sampling procedure or an alternative way to define our cross-modal loss may better suit the structural properties of our data.
Due to its simplicity, we focus on the latter option and test this hypothesis by introducing some modifications to Eq. (\ref{eq:nce}). By observing that similar captions are likely to correspond to similar audio, we can leverage this side information to estimate the \textit{relevance} between non-paired (negative) items in a mini-batch. Since more relevant items should contribute more to the learning process, similarly to \cite{Zolfaghari2021}, we can introduce a relevance-based weight:
\begin{equation} \label{eq:weight}
w_i = \exp \left(\frac{\frac{1}{N} \sum_j^N \operatorname{sim} (t_i, t_j)} {\kappa}\right),
\end{equation}
where $\kappa$ is a temperature hyperparameter and $\operatorname{sim} (t_i, t_j)$ a similarity score between text sample $t_i$ and text sample $t_j$. We can then redefine our audio-to-text loss in Eq. (\ref{eq:nce}) as follows:
\begin{equation}\label{eq:lw}
    \mathcal{L}^*_{a \rightarrow t}
    = - \frac{1}{N} \sum_{i}^{N} 
    w_i
    \log \frac{\exp \left(\boldsymbol{z}_{a, i} \cdot \boldsymbol{z}_{t, i}^{+} / \tau\right)}
    {\sum\limits_{\boldsymbol{z} \in\left\{\boldsymbol{z}_{t, i}^{+}, \boldsymbol{z}_{t, i}^{-}\right\}} 
    \exp \left(\boldsymbol{z}_{a, i} \cdot \boldsymbol{z} / \tau\right)}
\end{equation}
and obtain the total loss by summing this to its symmetrical counterpart $\mathcal{L}^*_{t \rightarrow a}$.
We dub this procedure \textit{content-aware loss weighting}, or \textit{loss weighting} for short.

\subsection{Combining MusCALL with Audio Self-Supervision}\label{sec:ssl}
At its core, the design of our approach is centred around audio-text matching. However, we are also interested in learning representations that can be transferred to other tasks, especially in a zero-shot way. Prior work has demonstrated the benefit of using self-supervised learning (SSL) \cite{Ericsson2021} to improve representation quality in a similar setting in the image domain \cite{Mu2021}. We hypothesise that our model may also benefit from this and experiment with combining our multimodal contrastive objective with self-supervision on the audio modality. Similarly to the approach proposed in \cite{Mu2021}, we do this via multi-task learning, using an adaptation of SimCLR \cite{Chen2020a} to the audio modality as the self-supervised component. Two correlated views of the audio input are produced via a data augmentation pipeline and passed through a shared audio encoder. The SSL objective is then computed on the embeddings resulting from the two views and added to our cross-modal objective as follows:
\begin{equation}\label{eq:ssl}
    \mathcal{L} = \lambda \mathcal{L}_{SSL} + (1 - \lambda) \mathcal{L}_{a \leftrightarrow t},
\end{equation}
where $\mathcal{L}_{SSL}$ is the \textit{NT-Xent} loss used in SimCLR \cite{Chen2020a} and $\lambda \in [0, 1]$ is a scalar weight.

We refer to the SSL-enhanced variant by \muscallssl{} to distinguish it from the base variant \muscallbase{}.

\input{tables/retrieval}

\subsection{Audio \& Text Encoding}
As our audio backbone, we choose ResNet-50 \cite{He2016} operating on melspectrogram representations of the input and adopt the same architectural modifications introduced in CLIP \cite{Radford2021}: 3 stem convolutions followed by average pooling instead of max pooling and anti-aliased blur pooling. We obtain fixed-length audio representations via an attention pooling mechanism: we append the average-pooled audio feature to the backbone output and compute multi-head self-attention, taking the output corresponding to the average-pooled feature as the global audio representation. 
Also analogously to CLIP, our text encoder is a Transformer \cite{Vaswani2017, Radford2019}. To avoid overfitting on our dataset, which is $\sim2$K times smaller than the dataset used to train CLIP, we downsize the network and use 4 hidden layers, as we empirically find this to be the optimal depth (see Figure \ref{fig:sensitivity}).

Two learned linear projections are applied to map the audio and text features produced by the audio and text backbones onto a $512$-dimensional multimodal embedding space respectively. The resulting features are then L2-normalised before their dot product is calculated in the contrastive loss.

\section{Experimental Design} \label{sec:experiments}
\subsection{Dataset}
We train and evaluate our model on a dataset of 250k (audio, text) pairs created from a production music library. This consists of full-length audio tracks, covering a broad range of genres, and a piece of text describing the overall musical content of each track, as shown in Table \ref{tab:examples}. 
For training, validation and testing, we use a random 80/10/10 split. 

\subsection{Implementation Details} \label{sec:details}
For each audio track, we take a $20s$ random crop at training time, and the central $20s$ segment at testing time. Unless otherwise specified, we then apply a stochastic data augmentation pipeline, where each transformation is applied with an independent probability $p$. We adopt some of the same transformations, such as noise injection and pitch shift, as prior work on music audio representation learning \cite{Spijkervet2021, Won2021}.
For each audio caption in the dataset, we take the text input in full and tokenise it following the same procedure as in CLIP, based on byte pair encoding \cite{Sennrich2016} with a 49K token vocabulary and maximum sequence length of 77.

As an estimate for text-text similarity in the calculation of our loss scaling weight $w_i$ in Eq. (\ref{eq:lw}), we use the cosine distance between L2-normalised embeddings produced by a pre-trained Sentence-BERT \cite{Reimers2019}. We set the loss weighting temperature parameter $\kappa$ to $0.005$, following \cite{Zolfaghari2021}.

For the SimCLR module in \muscallssl{}, following \cite{Spijkervet2021} we use a 256-dimensional non-linear projection layer and set the temperature parameter to 0.5. The loss scaling parameter $\lambda$ in Eq. (\ref{eq:ssl}) is set to 0.3 as we find this to yield best results.

Together with the parameters for the audio and text encoders and multimodal projections, we also learn the temperature parameter $\tau$ in Eq. (\ref{eq:nce}) to avoid tuning it as a hyperparameter.
We train using the Adam optimizer, with weight decay $0.2$, batch size of 256, initial learning rate of 5e-5, reduced throughout training following a cosine schedule. After training for a maximum of 150 epochs, we select the best model based on the R@$10$ score (see Section \ref{sec:retrieval}) computed on the validation set.
To save GPU memory, we perform training with automatic mixed precision.

\section{Experiments \& Results}\label{sec:results}
In this section we first describe our experimental setup and report our main results on cross-modal retrieval (Section \ref{sec:retrieval}), which constitute the focus of the paper. We then explore transferring our model to zero-shot music classification, highlighting key findings (Section \ref{sec:zeroshot_eval}).

\subsection{Cross-Modal Retrieval}\label{sec:retrieval}
In cross-modal retrieval, given a query item of modality $A$, our goal is to identify the matching item of modality $B$. We can do this by casting the retrieval as a ranking problem: for a given query item of modality $A$, we rank all candidate items of modality $B$ by the cosine similarity between the query embedding and each candidate embedding. The retrieval performance is then evaluated by computing the Recall at K (R@K) over the testing set as the percentage of correctly retrieved items within the top-K items for each query, with K $= \{1, 5, 10\}$. 
In line with previous work on cross-modal retrieval, we also report mean average Precision at 10 (mAP10) and Median Rank (MedR) in our main results. 
We construct our testing set by randomly sampling a subset of 1000 (audio, text) pairs from our testing split. This is chosen to be consistent in size with other datasets for sentence-based audio retrieval \cite{Drossos2020, Kim2019b}.

\paragraph*{Results}
Table \ref{tab:retrieval} shows the performance of \muscall{} on the cross-modal retrieval task. 
We compare this to the baseline system for the \textit{Language-Based Audio Retrieval} subtask of Task 6 in the 2022 DCASE Challenge,\footnote{\href{https://dcase.community/challenge2022/}{\texttt{https://dcase.community/challenge2022}}} trained and evaluated on our dataset. This is a simplified version of the approach proposed in \cite{Xie2022} and consists of a pre-trained \textit{word2vec} model \cite{Mikolov2013a} as the text encoder and a convolutional recurrent neural network as the audio encoder. Average pooling is used to obtain global representations for each modality from the output of the encoders. These are then jointly trained via a triplet ranking loss \cite{Bromley1993}. We also consider a variant of this baseline where we use our loss instead of the triplet ranking loss, to provide a closer comparison to our approach.

Our results show that MusCALL significantly outperforms the baseline on both text-to-audio and audio-to-text retrieval. Enhancing the baseline with our loss slightly reduces this margin, bringing a 14.2\% and 60.7\% average improvement over the vanilla baseline for text-to-audio and audio-to-text respectively. However, the use of our contrastive loss alone does not account for the full difference, indicating that the design of our text and audio encoders and the use of linear projection layers, play a crucial role.

\subsection{Zero-shot Transfer}\label{sec:zeroshot_eval}
\input{tables/zeroshot}

By design, our cross-modal learning approach endows the model with the ability to interpret arbitrary text inputs. This powerful property can be exploited to perform new tasks based on textual descriptions via a transfer learning paradigm known in the literature as \textit{zero-shot transfer} \cite{Radford2021, Jia2021, Zhai2021}. Similarly to zero-shot learning, zero-shot transfer aims to learn a model that can generalise to unseen classes or tasks without further training. But, in contrast to zero-shot learning, it somewhat relaxes the requirement that target classes must be completely unseen. Instead, the model is usually pre-trained in a task-agnostic fashion on large amounts of natural language text, and may be exposed to information relevant to the target tasks through this, although no supervised examples are provided.

We explore this paradigm in our work and investigate whether MusCALL exhibits zero-shot transfer capabilities on two tasks, genre classification and auto-tagging. We evaluate this on the most popular public datasets for these tasks, GTZAN \cite{Tzanetakis2002} and MagnaTagATune (MTAT) \cite{Law2009}. 
Treating audio clips in each dataset as queries, we obtain classification predictions by considering the similarity scores between audio embeddings and text embeddings of the target labels, similarly to the procedure described for cross-modal retrieval in Section \ref{sec:retrieval}. Based on this, we calculate accuracy for GTZAN, and area under the receiver operating characteristic curve (ROC-AUC) and area under the precision-recall curve (PR-AUC) for the MTAT dataset.

In order to reduce the distribution shift between pre-training and downstream text input when doing zero-shot transfer, it is common practice to wrap the labels in templates. For example, the label ``\textit{rock}'' may be wrapped in the sentence ``\textit{This is a rock song with electric guitars}'', making it much closer to a typical caption encountered in training. Based on evidence that it can improve performance \cite{Radford2021}, we explore this in our evaluation by passing ``\textit{A} [\textsc{label}] \textit{track}'' as the text input, but do not further tune the prompts to our model, leaving this for future work. 

\begin{figure}[t]
     \centering
     \begin{subfigure}[b]{0.23\textwidth}
         \centering
         \includegraphics[width=\textwidth]{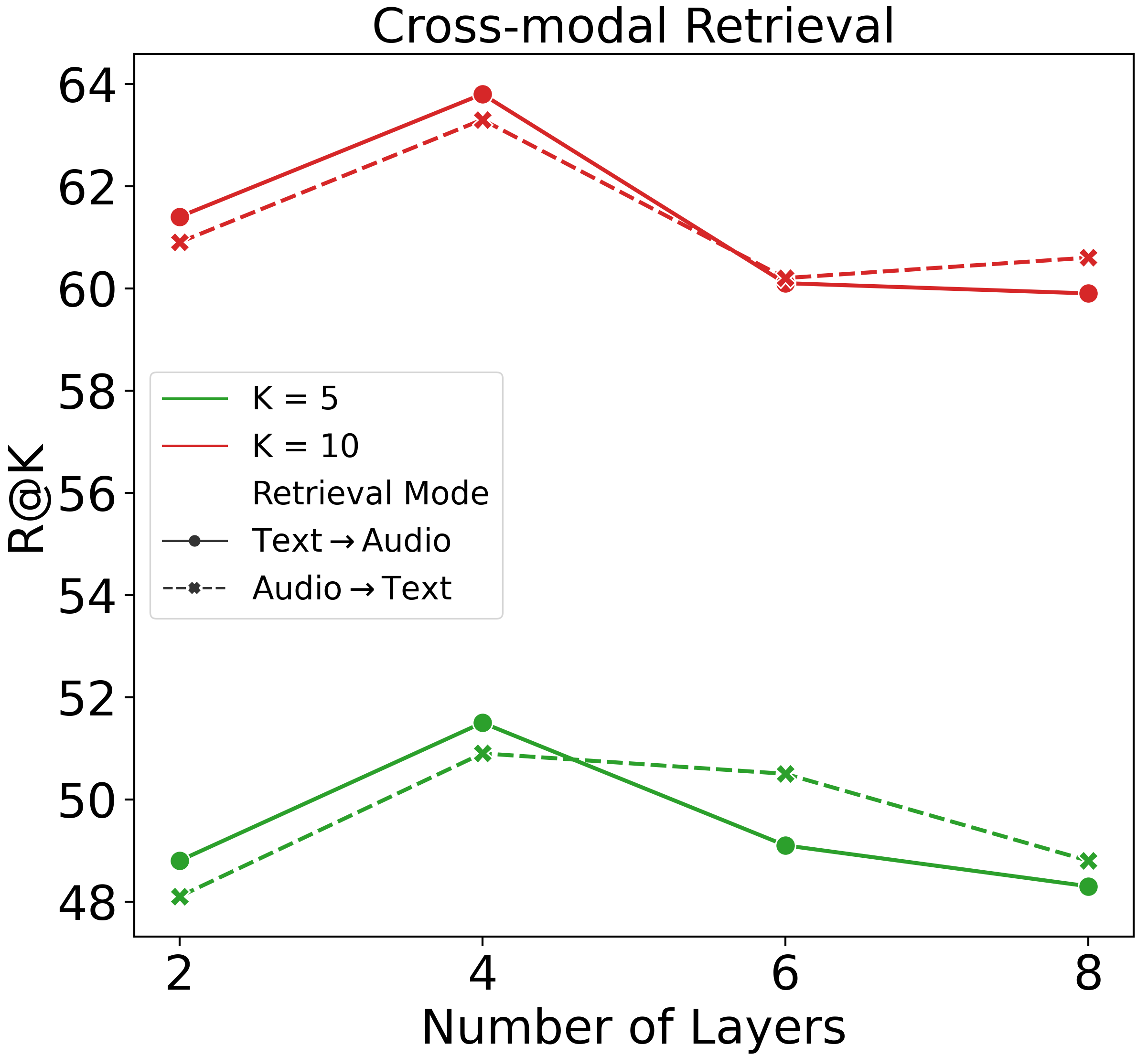}
     \end{subfigure}
     \hfill
     \begin{subfigure}[b]{0.23\textwidth}
         \centering
         \includegraphics[width=\textwidth]{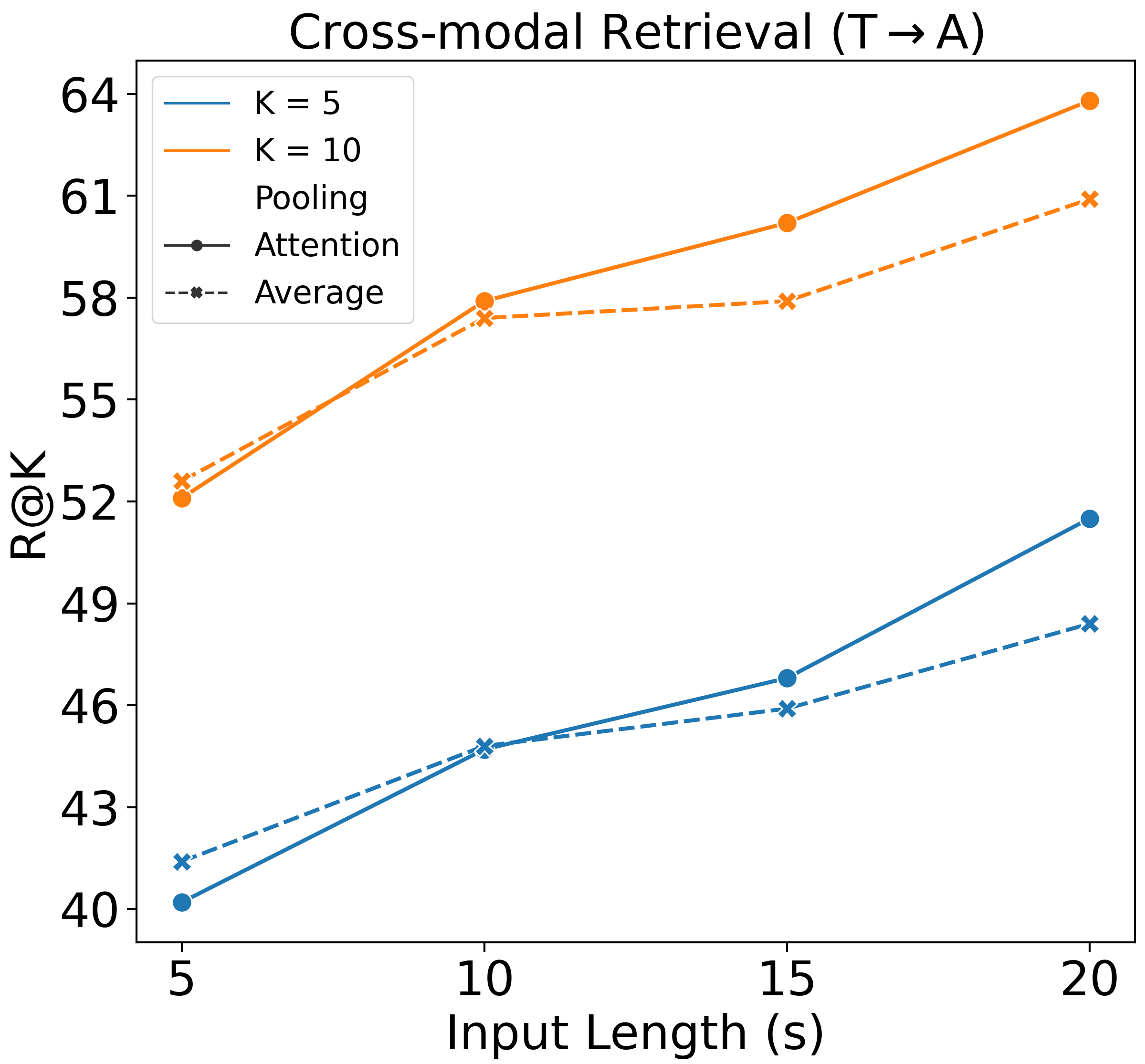}
     \end{subfigure}
        \caption{\textbf{The effect of varying} (left) the \textbf{depth} of the text encoder and (right) the \textbf{audio input length} on retrieval.}
        \label{fig:sensitivity}
        \vspace{-0.4cm}
\end{figure}

\paragraph*{Results}
In Table \ref{tab:zeroshot} we compare zero-shot performance of two model variants, \muscallbase{} and \muscallssl{}.
We find that the self-supervised learning objective yields, on average, better performance, with a $5.4\%$ improvement over \muscallbase{}. This is not too surprising, as the SSL objective is designed to improve the representation quality of our audio branch and is therefore expected to have a positive effect on generalisation \cite{Wang2022}. However, this improvement is not consistent across datasets and tasks, a result that may be attributed to different degrees of similarity between pre-training and downstream datasets, as found in prior work \cite{Hendricks2021}. We also find that zero-shot performance is sensitive to the choice of text prompt. This confirms a well-known phenomenon reported in the literature \cite{Radford2021, Lester2021} and suggests that techniques such as prompt tuning and ensembling \cite{Radford2021} may lead to further improvements. 

Additionally, there are some important differences that should be considered when comparing \muscallbase{} and \muscallssl{}, since the SSL module introduces more activations and overall learnable parameters. This has two implications: firstly, to offset the increased memory footprint while keeping the batch size unchanged and still satisfying our memory constraints, we reduce the size of the input audio to $10s$. We verify that this slightly degrades performance on cross-modal retrieval even in the absence of the SSL module, as shown in Figure \ref{fig:sensitivity}. Assuming that this trend would be reflected in the zero-shot scenario, we conclude that \muscallssl{} would benefit from using longer audio clips. Secondly, the higher number of parameters makes \muscallssl{} prone to overfitting, a factor that we believe may also be limiting its performance. 

\section{Analysis \& Discussion} \label{sec:discussion}

We now discuss the qualitative characteristics of our model (Section \ref{sec:qualitative}) and examine the contributions of the main design choices through an ablation study (Section \ref{sec:ablations}).

\subsection{Qualitative Analysis}\label{sec:qualitative}

\begin{figure}
     \centering
     \begin{subfigure}[b]{0.23\textwidth}
         \centering
         \includegraphics[width=\textwidth]{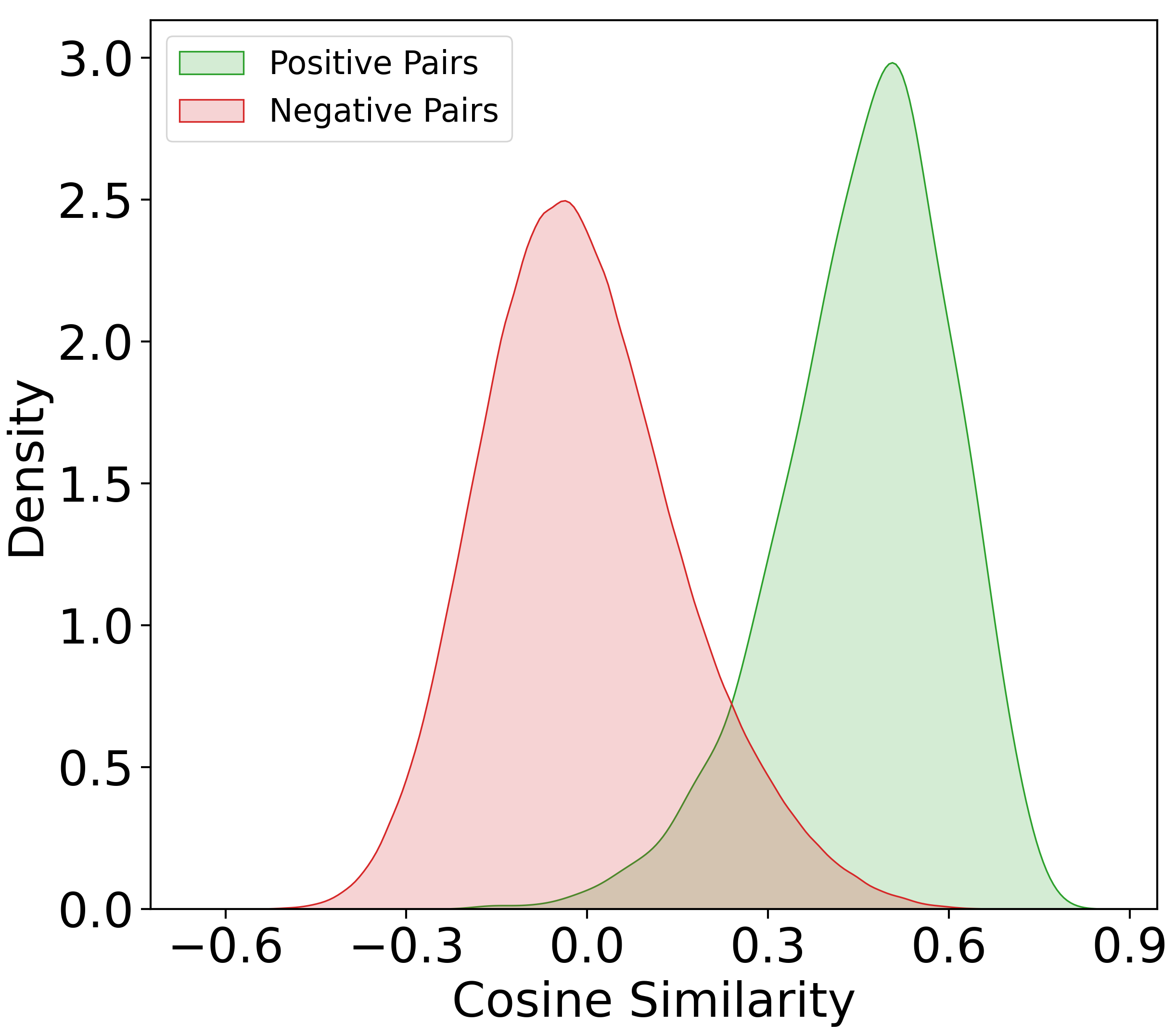}
     \end{subfigure}
     \hfill
     \begin{subfigure}[b]{0.24\textwidth}
         \centering
         \includegraphics[width=\textwidth]{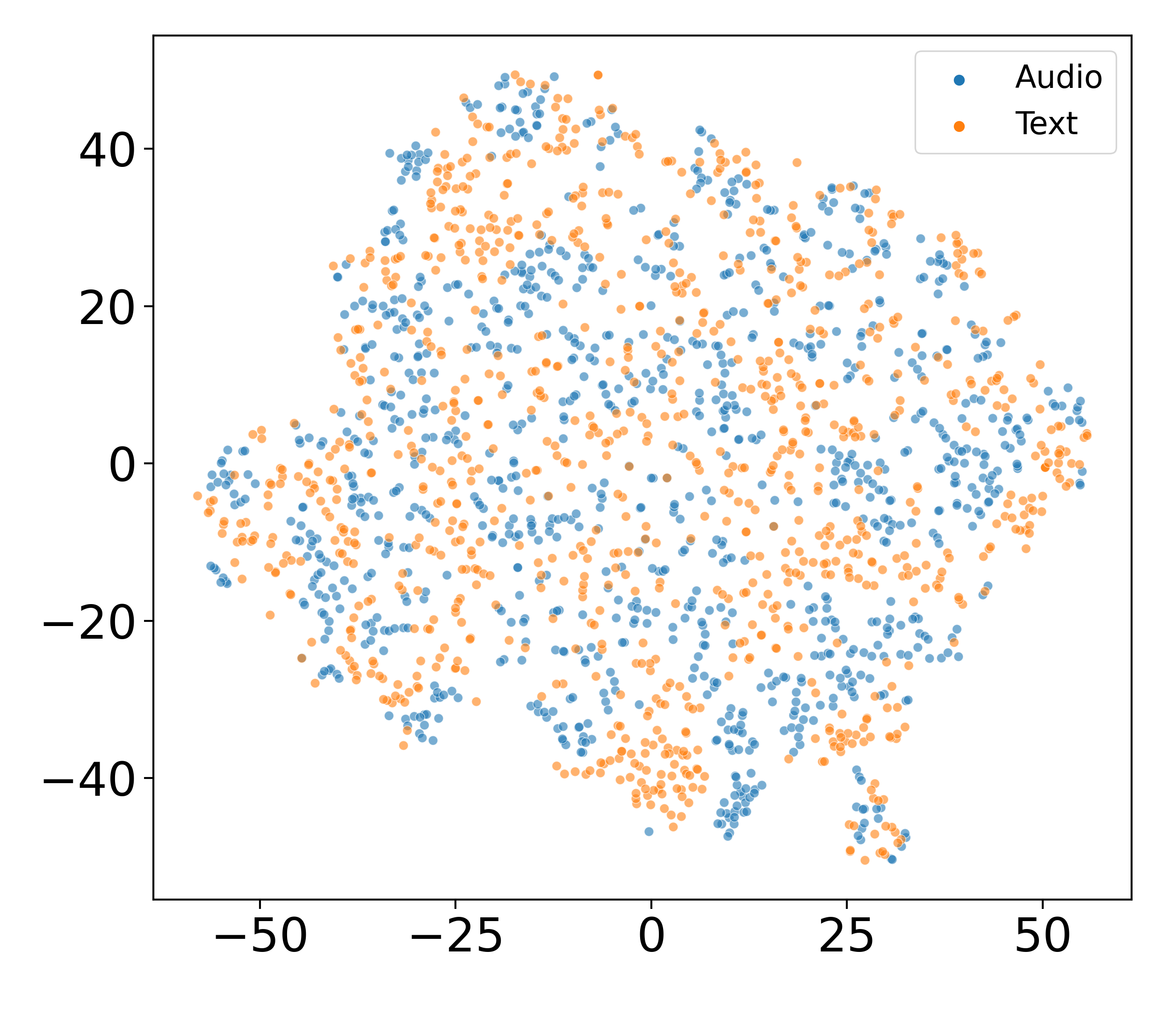}
     \end{subfigure}
       \caption{\textbf{Feature visualisation}. (Left) distributions of pair-wise similarity scores of aligned and non-aligned (audio, text) pairs in our testing set. (Right) t-SNE visualisation of audio and text features in the multimodal output space.}
        \label{fig:feat_visualization}
        \vspace{-0.4cm}
\end{figure}

\paragraph*{Similarity score distribution}
In Figure \ref{fig:feat_visualization} (left) we show the kernel density estimation of the distributions of pairwise similarity scores in the joint embedding space for positive and negative pairs in our testing set. From this we can see that aligned pairs have distinctly higher scores compared to random ones, confirming that the model distinguishes positive and negative examples with a good level of confidence.

\paragraph*{Feature visualisation}
Figure \ref{fig:feat_visualization} (right) shows a t-SNE plot of audio and text embeddings in our testing set. We observe no separation between modalities in the output space, with representations of both audio and text inputs well mixed together, confirming that MusCALL learns to adequately map both input modalities to a common space.

\paragraph*{Failure analysis}
In order to provide more context to our cross-modal retrieval results beyond metric-based evaluation, we also perform a qualitative error analysis by examining the relevance of the retrieved items in cases of incorrect retrieval. As highlighted in the examples in Table \ref{tab:examples}, in
some cases, while the ground-truth item is not ranked first,
MusCALL is still capable of retrieving audio tracks whose
associated caption is semantically related to the query.

\subsection{Ablation Study}\label{sec:ablations}
\paragraph*{The effect of data augmentations \& random crop}
Table \ref{tab:ablations} shows that removing random cropping (RC) considerably degrades performance (-13.7\% compared to \muscallbase{}), while removing the audio augmentations (AA) only marginally alters results (-1.5\%). When considering both together, we find that, in the absence of RC, the benefit of performing augmentations is amplified, and removing both has more drastic effect (-31.5\%). This is expected, since both techniques effectively increase the variety of samples seen in training. To avoid costly hyperparameter tuning, we do not exhaustively explore audio transformations and their composition, and note that tailoring the AA pipeline, for example via additional frequency-domain transformations, may lead to better results.

\paragraph*{The effect of loss weighting}
As also shown in Table \ref{tab:ablations}, using our content-aware loss weighting (Section \ref{sec:lw}) results in comparable retrieval performance to our vanilla contrastive loss. To more closely observe the effect of using loss weighting, we compare the pairwise similarity distributions of positive and negative pairs produced by MusCALL with and without loss weighting. This reveals that similarity scores of positive pairs have lower variance and a higher mean when using loss weighting, suggesting that this technique nudges the learning process towards a better discrimination of positives and negatives, but not enough to produce significant improvements in the cross-modal retrieval task.

\input{tables/examples}
\input{tables/ablations}

\paragraph*{The effect of attention pooling}
Our ablation study confirms that removing attention pooling hurts performance: on average, the Recall results drop by $4.9\%$ compared to simple average pooling. We note that the advantage of using attention pooling becomes more pronounced as we increase the audio input length, as shown in Figure \ref{fig:sensitivity}. This is particularly relevant in the music domain, since music signals exhibit structure over longer ranges compared to other types of audio signals like environmental sounds.

\section{Conclusion}
We presented MusCALL, a method for multimodal contrastive learning of audio-linguistic representations. By leveraging aligned (audio, text) pairs, MusCALL successfully learns to perform cross-modal retrieval, allowing to search for music via natural language queries and vice versa. Extending this multimodal alignment capability to the zero-shot setting, MusCALL can also be transferred to music classification tasks by simply providing target labels as text inputs. Through an extensive set of experiments, we validated the main design choices in our core approach and explored two variants. Through the first variant, we demonstrated the viability of using a text-based similarity metric to weigh the loss contribution of each negative sample, providing a starting point for improving multimodal contrastive learning on real-world music data. Through the second variant, we explored including a self-supervised objective to improve the audio representation quality. Both variants show promising results and provide opportunities for further research. Future work will focus on assessing their performance in more depth, particularly in the zero-shot scenario, including human evaluations alongside automatic metrics and considering a wider set of tasks.

\section{Acknowledgements}
This work was supported by UK Research and
Innovation [grant number EP/S022694/1] and Universal Music Group. We would also like to thank The Alan Turing Institute for the support provided during the first author's time as an Enrichment Student.

\bibliography{muscall}

\end{document}

%% file: tables/retrieval.tex
\begin{table*}[t]
    \centering
    \begin{tabular}{lcccccccccc}
        \multirow{2}{*}{Method} &
        \multicolumn{5}{c}{Text $\rightarrow$ Audio} & \multicolumn{5}{c}{Audio $\rightarrow$ Text} \\
        \cmidrule(rl){2-6}  \cmidrule(rl){7-11}
        & R@1   & R@5   & R@10  & mAP10   & MedR $\downarrow$ & R@1  & R@5   & R@10  & mAP10   & MedR $\downarrow$ \\ \cmidrule(rl){1-11}
        DCASE \cite{Xie2022}              &    2.3   &    10.4   &   17.4 & 5.5 &  50  &   1.1    &   5.6  &  10.1   & 3.0 & 84    \\ 
        DCASE + CL   &    3.9   &    12.4   &    18.1 & 6.8 & 81.5  &   2.0    &   8.6  &    16.4   & 4.5 &  64 \\ 
     \cdashlinelr{1-11}
        MusCALL (ours)        &  \textbf{25.9}    &    \textbf{51.9}   &    \textbf{63.3}  & \textbf{36.0} & \textbf{5} & \textbf{25.8} &  \textbf{53.0}  &  \textbf{63.0}  & \textbf{35.9} & \textbf{5} \\  
        \cmidrule(rl){1-11}
    \end{tabular}
        \caption{\textbf{Cross-modal retrieval results}. MusCALL improves performance on all metrics by a large margin, compared to two variants of the baseline: one with the original triplet ranking loss (DCASE) and one with our loss (DCASE + CL).}
    \vspace{-0.3cm}
    \label{tab:retrieval}
\end{table*}

%% file: tables/zeroshot.tex
\begin{table}[t]
    \centering
    \begin{tabular}{lcccc}
    \toprule
         \multirow{2}{*}{Method} & \multirow{2}{*}{Prompt} & Genre & \multicolumn{2}{c}{Tagging} \\
       \cmidrule(rl){3-3} \cmidrule(rl){4-5}
       & & Acc. & ROC & PR       \\
       \midrule
     \muscallbase{} & \xmark &  55.5  &    \textbf{78.0} & 28.3   \\ 
       \muscallbase{} & \cmark & 52.0   &   72.0  & 21.0   \\ 
        \cdashlinelr{1-5}
         \muscallssl & \xmark &    58.2    & \textbf{77.4} & \textbf{29.3}  \\   
         \muscallssl & \cmark &    \textbf{62.0}    &  73.4 & 23.2 \\ 
         \bottomrule
    \end{tabular}
    \caption{\textbf{Zero-shot transfer results}. We report \muscallbase{} and \muscallssl{} accuracy on GTZAN (genre classification), and ROC-AUC and PR-AUC on MTAT (auto-tagging). \textit{Prompt} indicates whether a template was used to wrap the class label.}
    \vspace{-0.3cm}
    \label{tab:zeroshot}
\end{table}

%% file: tables/examples.tex
\begin{table}[t]
\centering
    \small
    \begin{tabular}{p{3.6cm}p{3.6cm}}
        Query Text & Text of the Top-1 Audio \\
        \cmidrule(rl){1-2} 
        \rowcolor{gray!10}
        \textit{An atmospheric and introspective orchestral track featuring strings, piano, and synth.} & \textit{An inspirational and moody orchestral track featuring strings and choir.} \\
        \textit{Deep chilled out space jazz with crisp beats and lush electronics.} &
        \textit{Jaunty swing featuring trumpet.} \\
        \rowcolor{gray!10}
        \textit{Up tempo, pumping dance pop with female vocals.} &
        \textit{Quirky, fun, positive disco party music.} \\
        \cmidrule(rl){1-2} 
    \end{tabular}
    \caption{\textbf{Failure analysis} of text-to-audio retrieval.}
    \vspace{-0.2cm}
    \label{tab:examples}
\end{table}

%% file: tables/ablations.tex
\begin{table}[t]
\centering
    \begin{tabular}{llllccc}
        \multirow{2}{0.7cm}{LW} &
        \multirow{2}{0.7cm}{RC} &
        \multirow{2}{0.7cm}{AA} &
        \multirow{2}{0.7cm}{AP} &
        \multicolumn{3}{c}{Text $\rightarrow$ Audio} \\
        \cmidrule(rl){5-7} 
        &    &   &  & R@1   & R@5   & R@10   \\ \cmidrule(rl){1-7}
        \xmark & \xmark & \xmark &  \cmark &  14.7    &    33.9   &      47.2        \\ 
        \xmark & \xmark & \cmark & \cmark &  18.5    &    43.5  &     57.4       \\ 
        \xmark & \cmark &  \cmark &  \xmark  &   24.9  &    49.3  &   58.9     \\ 
        \xmark & \cmark & \xmark & \cmark &   24.3    &   51.3    &      62.2 \\
        \xmark &  \cmark &  \cmark &  \cmark &   24.6    &    51.5   &    63.8 \\
        \cdashlinelr{1-7}
        \cmark & \cmark & \cmark & \cmark &   25.9    &    51.9   &    63.3 \\
        \cmidrule(rl){1-7}
    \end{tabular}
    \caption{\textbf{Ablations}: loss weighting (LW), random cropping (RC), audio augmentations (AA), attention pooling (AP).}
     \vspace{-0.3cm}
    \label{tab:ablations}
\end{table}